\documentclass[aps,prd,preprint,groupedaddress]{revtex4}

\usepackage{graphicx}

\def\vc#1{\textbf{\textit #1}}

\def\varPhi{\mathit{\Phi}}
\def\varLambda{\mathit{\Lambda}}
\def\r2{\sqrt 2}
\def\h#1{h_#1^0}
\def\M#1#2{{\cal M}_{#1#2}^0}
\def\O#1i{O_{#1i}}
\def\th#1{\theta_#1}
\def\ta#1{\alpha_#1}
\def\PL{\frac{1-\gamma_5}{2}}
\def\PR{\frac{1+\gamma_5}{2}}
\def\Hi{\tilde H_i^0}
\def\MH#1{\tilde M_{H#1}^2}
\def\sigH#1{\sigma(\tilde H_#1^0)}

\def\sigSMH#1{\sigma_{SM}(\tilde H_#1^0)}

\def\tanb{\tan\beta}

\def\PRev#1#2#3{Phys. Rev. {#1}, #2 (#3)}
\def\PRD#1#2#3{Phys. Rev. {\bf D #1}, #2 (#3)}
\def\NPB#1#2#3{Nucl. Phys. {\bf B #1}, #2 (#3)}
\def\PTP#1#2#3{Prog. Theor. Phys. {\bf #1}, #2 (#3)}
\def\PTEP#1#2#3{Prog. Theor. Exp. Phys. {\bf #1}, #2 (#3)}

\def\EPJC#1#2#3{Eur. Phys. J. {\bf C #1}, #2 (#3)}
\def\PLB#1#2#3{Phys. Lett. {\bf B #1}, #2 (#3)}
\def\PRL#1#2#3{Phys. Rev. Lett. {\bf #1}, #2 (#3)}
\def\PRep#1#2#3{Phys. Rep. {\bf #1}, #2 (#3)}

\def\CPC#1#2#3{Chin. Phys. {\bf C #1}, #2 (#3)}
\def\JHEP#1#2#3{JHEP {\bf #1}, #2 (#3)}

\begin{document}

\preprint{
OCHA-PP-337
}

\title{
Higgs bosons in supersymmetric model \\
with CP-violating potential
}


\author{
Noriyuki Oshimo
}
%\email[]{oshimo@phys.ocha.ac.jp}
%\thanks{}
\affiliation{
Department of Physics,
Ochanomizu University,
Tokyo, 112-8610, Japan
}


\date{\today}

\begin{abstract}

     In the supersymmetric standard model which is not minimal, the Higgs potential 
does not conserve CP symmetry generally.   
Assuming that there exists an SU(2)-triplet Higgs field, 
we discuss resultant CP-violating effects on the Higgs bosons.    
The experimentally observed Higgs boson, which should be CP-even in the standard model, 
could decay into two photons of CP-odd polarization state non-negligibly.  
For the second lightest Higgs boson, in sizable region of parameter space, 
the dominant decay modes are different from those expected by
the standard model.  
The two-photon decay could yield both even and odd CP final states at a ratio of oder of unity.  

\end{abstract}

%\keywords{}

\maketitle

%\section{}
%\subsection{}
%\subsubsection{}

\section{Introduction\label{intro} } 

     Now that the last missing piece of the standard model (SM), 
the Higgs boson, has been discovered \cite{exp}, it would become 
a main subject in particle physics to pursue theory beyond the SM.    
However, there are not many phenomena observed 
which can provide clues to the new theory.  
Examining the Higgs boson and its related phenomena from various aspects is 
thus very important.  
In experimental measurements for the Higgs boson,  the production cross section \cite{expbb} 
and the branching ratios of the decays into $WW^*$ \cite{expww}, $ZZ^*$ \cite{expzz}, 
and $\gamma\gamma$ \cite{exprr} are not inconsistent with the SM.  
However, there seems to be sizable room for allowing theory which deviates from it.  
The awaited clues may possibly be found in studying the Higgs boson.  
   
     From a theoretical point of view, supersymmetry may be considered 
one fundamental symmetry existing in nature.  
If this conjecture is true, the Higgs sector of the resultant model becomes
very different from the SM.  
Even in the minimal supersymmetric extension of the SM (MSSM), 
there exist three neutral Higgs bosons and a charged Higgs boson.  
More involved extensions have more rich Higgs sectors.  
These differences for the Higgs sector could become a informative guide for theory 
beyond the SM, and thus would be worth studying.  
In particular, squarks and sleptons, typical particles evidential of supersymmetry, 
may be inaccessible in near future experiments, owing to their 
possible large masses \cite{oshimo1}.  
Supersymmetry might be examined only in the Higgs sector or indirect effects 
such as CP violation \cite{oshimo2}.  

     We study violation of CP symmetry for the Higgs bosons of the supersymmetric model 
in which the Higgs sector contains an SU(2)-triplet superfield besides 
ordinary two doublet ones.  
This CP symmetry is conserved in the Higgs sector at tree level within the framework of 
the SM or of the MSSM.  
However, the additional superfield violates CP invariance generally at tree level.  
All the complex coefficients of the Lagrangian cannot be eliminated by 
redefining particle fields, unless some accidental cancellation is assumed.  
Consequently, the Higgs bosons in mass eigenstates become linear  
combinations of CP-even and CP-odd scalar fields.  
The mass spectrum is not determined trivially by magnitudes alone of the coefficients.  
Conservation of CP symmetry is no more respected in the interactions of the Higgs bosons 
with the quarks and leptons.  
Such a triplet superfield could be predicted by the SU(5) grand unified theory 
which contains a 24 dimensional Higgs boson.   

     One possible effect of CP violation for the Higgs bosons could appear 
in the decay into two photons \cite{oshimo3}.  
This decay process is generated at one-loop level, to which the $t$ quark, 
$b$ quark, or $W$ boson contributes dominantly.  
Since the Higgs boson is not in a CP eigenstate, 
both CP-even and CP-odd final states are induced for polarization of the two photons.  
This CP-violating effect can be observed even if the supersymmetric $R$-odd particles are all 
heavy and undetectable in near future experiments.  
We discuss disagreement of CP eigenstate and mass eigenstate for the Higgs bosons, 
and perform numerical analyses of the widths for polarization states which could be detected 
by measuring the polarization planes of the photons.  

     In the MSSM it may be possible that radiative corrections by the interactions 
with the $t$ squarks generate mixing of CP-even and CP-odd fields 
for the Higgs boson \cite{pilaftsis}.  
However, experimental upper bounds on the electric dipole moment of the neutron 
tells that the $u$ and $d$ squark masses should be at least of order of a few TeV if CP-violating phases 
of relevant coefficients are not suppressed.  
Assuming that the squark masses do not depend much on their generations, 
the possible mixing of CP eigenstates becomes small.  
The supersymmetric model with the additional triplet Higgs field has been 
studied in the literature \cite{chiara}, though CP invariance is assumed by taking relevant 
coefficients real.  
Non-conservation of CP symmetry yields different aspects for phenomenological features.  

     In sect.\ref{model} our model is briefly described.  In sect.\ref{int},  
calculating mass eigenstates for the Higgs bosons, we obtain their 
interactions with the quarks and gauge bosons.  
In sect.\ref{decay} we discuss the polarization CP eigenstates for the two-photon decay of 
the Higgs boson.  
The mixing of CP-even and CP-odd final states is studied numerically in sect.\ref{analyses}.  
Discussions are given in sect.\ref{discussions}.    

\section{Model\label{model}} 
 
      The supersymmetric standard model may contain, under grand unified theory, 
a Higgs superfield of SU(2) triplet.  
In the SU(5) model, the Higgs sector consists of the superfields 
belonging to ${\bf 5}$, ${\bf\bar 5}$, and ${\bf 24}$ representations, being denoted 
respectively by $H$, $\bar H$, and $\varPhi_5$.  
The superpotential is given by 
\begin{equation}
  W = M_H{\bar H}H + \frac{1}{2}M_{\varPhi 5}{\rm Tr}[\varPhi_5^2]
       + \frac{1}{3}\lambda_\varPhi{\rm Tr}[\varPhi_5^3] +  \lambda_{H\varPhi} {\bar H}\varPhi_5 H. 
\end{equation} 
After spontaneous breaking of SU(5) gauge symmetry, it could happen that the SU(3)-singlet 
component of $\varPhi_5$ receives a mass much smaller than the grand unification scale, 
as $H$ and $\bar H$ should do.  
Then, the Higgs sector of the supersymmetric standard model is described by the superpotential 
\begin{equation}
    W = -\mu_H H_1\epsilon H_2 
          + \frac{1}{2}\mu_\varPhi{\rm Tr}[\varPhi^2]
      -  \lambda H_1\epsilon\varPhi H_2 ,
\end{equation}
where $H_1$, $H_2$, and $\varPhi$ are transformed as  $({\bf 1},{\bf 2},-1/2)$, 
$({\bf 1},{\bf 2},1/2)$, and $({\bf1},{\bf3}, 0)$ under SU(3)$\times$SU(2)$\times$U(1) gauge symmetry, 
respectively, with $\epsilon$ being the antisymmetric tensor of rank 2.  
The dimensionless parameter $\lambda$  and the mass parameters $\mu_H$ and $\mu_\varPhi$ 
have all complex values generally.   
The supersymmetry soft-breaking terms of the Lagrangian density are expressed as 
\begin{eqnarray}
    \cal L_{SB} &=& m_1^2 H_1\epsilon H_2 
          - \frac{1}{2}m_2^2{\rm Tr}[\varPhi^2]
       + m_3 H_1\epsilon\varPhi H_2 +{\rm H.c.}  
   \nonumber \\
   & & - {\rm Re}(M_{H1}^2)|H_1|^2-{\rm Re}(M_{H2}^2)|H_2|^2
   -{\rm Re}(M_\Phi^2){\rm Tr}[\varPhi^\dagger\varPhi],
\end{eqnarray}
where the scalar fields are denoted by the same symbols as 
their corresponding superfields.   
The mass parameters $m_1$, $m_2$, and $m_3$ have generally complex values.  
The parameters $M_{H1}^2$, $M_{H2}^2$, and $M_\Phi^2$ have mass-squared dimension.  
We write the contents of the Higgs fields $\varPhi$, $H_1$, and $H_2$ as 
\begin{eqnarray}      
    \varPhi &=& \frac{1}{\r2}\left(
     \matrix{ \phi^0     & \r2\phi^+ \cr
                   \r2\phi^-    &    -\phi^0}       
           \right), 
\end{eqnarray}

\begin{equation}
   H_1 = \left(
      \matrix{\h1 \cr
                   h_1^-}
      \right), \quad \quad
   H_2 = \left(
      \matrix{h_2^{+}\cr
                   \h2}
      \right).  
\end{equation}
The neutral components are expressed by  
\begin{equation}
      \phi^0 = \frac{1}{\r2}(\phi_R+i\phi_I), 
\end{equation}
\begin{equation}      
    \h1 = \frac{1}{\r2}(h_R^1+ih_I^1), \quad \quad \h2 = \frac{1}{\r2}(h_R^2+ih_I^2),    
\end{equation}
where the fields with index $R$ or $I$ are real scalar bosons.  

     The terms of the scalar potential at tree level which consist only of the neutral components 
are given by 
\begin{eqnarray}
   V_0 &=& M_1^2|\h1|^2 + M_2^2|\h2|^2 + M_3^2|\phi^0|^2 
   \nonumber \\
   &+& r_1(|\h1|^4+|\h2|^4) + r_2|\h1|^2|\h2|^2 + r_3(|\h1|^2+|\h2|^2)|\phi^0|^2 
    \nonumber \\
    &+& \{-m_1^2\h1\h2 + \frac{1}{2}m_2^2\phi^0\phi^0 -\frac{1}{\r2}\lambda \mu_H^*(|\h1|^2+|\h2|^2)\phi^0 
    \nonumber  \\
           & & + \frac{1}{\r2}\lambda \mu_\phi^*\phi^{0*}\h1\h2 + \frac{1}{\r2}m_3\phi^0\h1\h2 + {\rm H.c.} \},  
\label{treepot}
\end{eqnarray}
where the coefficients are defined as 
\[
  M_1^2 = |\mu_H|^2 + {\rm Re}(M_{H1}^2), \quad 
  M_2^2 = |\mu_H|^2 + {\rm Re}(M_{H2}^2), \quad 
  M_3^2 = |\mu_\phi|^2 + {\rm Re}(M_\Phi^2), 
\]
\begin{equation}
  r_1 = \frac{1}{8}(g^2+g'^{2}), \quad 
  r_2 = -\frac{1}{4}(g^2+g'^2) + \frac{1}{2}|\lambda|^2, \quad 
  r_3 = \frac{1}{2}|\lambda|^2.  
\end{equation}
Here, $g$ and $g'$ stand for the gauge coupling constants for SU(2) and U(1), respectively.  
The mass-squared parameters $M_i^2$ ($i$=1-3) and the dimensionless 
parameters $r_i$ ($i$=1-3) have real values.    
 
     Besides the tree-level terms, the scalar potential receives sizable contributions from 
radiative corrections.   Among them, the dominant contribution 
is mediated by the $t$ quark and $t$ squarks \cite{okada}.  
Denoting the $t$-quark mass and $t$-squark masses by $m_t$ and $M_{ti}$ ($i=$1,2), 
respectively, the correction terms at one-loop level is given by 
\begin{eqnarray}
  V_1 &=& -\frac{3}{16\pi^2}m_t^4\left(\log\frac{m_t^2}{\varLambda^2}+\frac{1}{2}\right)  
        \nonumber \\
     &+& \frac{3}{32\pi^2}M_{t1}^4\left(\log\frac{M_{t1}^2}{\varLambda^2}+\frac{1}{2}\right) 
     + \frac{3}{32\pi^2}M_{t2}^4\left(\log\frac{M_{t2}^2}{\varLambda^2}+\frac{1}{2}\right), 
 \label{onepot}
\end{eqnarray}
where $\varLambda$ is an appropriate energy scale.  
The $t$ quark receives a mass from the vacuum expectation value (VEV) of the Higgs boson,  
\begin{equation}
m_t ^2= |\eta_t\langle\h2\rangle|^2, 
\end{equation}
with $\eta_t$ being a coupling constant.  
For simple and definite calculations, we approximate the $t$-squark masses at 
\begin{equation}
   M_{t1}^2 = |\eta_t\langle\h2\rangle|^2 + {\rm Re}(M_Q^2), 
   \quad M_{t2}^2 = |\eta_t\langle\h2\rangle|^2 + {\rm Re}(M_{U^c}^2), 
 \label{stop}
\end{equation}
where $M_Q^2$ and $M_{U^c}^2$ stand for mass-squared parameters arising from supersymmetry 
soft-breaking terms.  
The energy scale $\varLambda$ is taken as 
\begin{equation}
      \langle -2m_t^2\left(\log\frac{m_t^2}{\varLambda^2}+1\right)  
     + M_{t1}^2\left(\log\frac{M_{t1}^2}{\varLambda^2}+1\right) 
     + M_{t2}^2\left(\log\frac{M_{t2}^2}{\varLambda^2}+1\right)\rangle = 0 
     \label{scale} 
\end{equation} 
for convenience.

\section{Interactions\label{int}} 

     The mass eigenstates for Higgs bosons, which participate in interactions with other 
particles, are determined by the scalar potential $V$.  
The potential is given by the sum of the tree level potential in Eq. (\ref{treepot}) 
and one-loop corrections in Eq. (\ref{onepot}), $V=V_0+V_1$.  
In general, this potential has five complex parameters  $\lambda \mu_H^*$, $\lambda \mu_\phi^*$, 
and $m_i$ ($i$=1-3).  
Although two coefficients can be made real by redefining phases of the fields,  
three coefficients remain complex, leading to CP violation.  
Taking $m_1^2$ and $m_2^2$ for real without loss of generality,  
we define $m_1^2=|m_1^2|$, $m_2^2=-|m_2^2|$, 
$\lambda \mu_H^*=|\lambda \mu_H^*|{\rm e}^{i\ta1}$, 
$\lambda \mu_\phi^*=|\lambda \mu_\phi^*|{\rm e}^{i\ta2}$, and 
$m_3=|m_3|{\rm e}^{i\ta3}$.  
  
     Owing to the complex coefficients, the VEVs of the Higgs bosons become complex.  
Assuming that electromagnetic symmetry is not broken, we express the VEVs as 
\begin{equation}
   \langle\h1\rangle = v_1{\rm e}^{i\th1}, \quad 
   \langle\h2\rangle = v_2{\rm e}^{i\th2}, \quad 
   \langle\phi^0\rangle = v_0{\rm e}^{i\th0}, 
\end{equation}
where $v_1$, $v_2$, and $v_0$ are the absolute values.  
The masses of the $Z$ and $W$ bosons then become 
\begin{eqnarray}
 M_Z &=& \frac{1}{\r2}\sqrt{(g^2+g'^2)(v_1^2+v_2^2)}, 
 \label{Zmass}  \\
 M_W &=& \frac{1}{\r2}g\sqrt{v_1^2+v_2^2+4v_0^2},  
 \label{Wmass}
\end{eqnarray}
so that the $\rho$ parameter is given by 
\begin{equation}
 \rho = \frac{v_1^2+v_2^2+4v_0^2}{v_1^2+v_2^2}.
\end{equation}
Since the complex phases $\th1$ and $\th2$ appear as 
a linear combination $\th1+\th2$ $(\equiv \theta)$ in the potential, 
only this combination is determined at the vacuum in our scheme.  
The extremum conditions of the potential $V$ for 
$v_1$, $v_2$, $v_0$, $\theta$, and $\th0$ are respectively given by 
\begin{eqnarray}
   && v_1\left\{M_1^2 + 2r_1v_1^2 + r_2v_2^2 + r_3v_0^2 - 
                                         \r2 v_0|\lambda \mu_H^*|\cos(\ta1+\th0)\right\} = 
      \nonumber \\
   &&   \ \    v_2\left\{|m_1^2|\cos\theta  - \frac{1}{\r2}v_0|\lambda \mu_\phi^*|\cos(\ta2-\th0+\theta) 
                                              - \frac{1}{\r2}v_0|m_3|\cos(\ta3+\th0+\theta)\right\}, 
       \label{extremum1}\\
   && v_2\left\{M_2^2 + 2r_1v_2^2 + r_2v_1^2 + r_3v_0^2 - 
                                        \r2 v_0|\lambda\mu_H^*|\cos(\ta1+\th0)\right\} = 
      \nonumber \\
   &&   \ \    v_1\left\{|m_1^2|\cos\theta  - \frac{1}{\r2}v_0|\lambda \mu_\phi^*| \cos(\ta2-\th0+\theta)
                                                - \frac{1}{\r2}v_0|m_3| \cos(\ta3+\th0+\theta)\right\}, 
 \label{extremum2} \\
  && v_0\left\{M_3^2 + r_3(v_1^2+v_2^2) -|m_2^2|\cos2\th0 \right\} 
                                              - \frac{1}{\r2}(v_1^2+v_2^2)|\lambda\mu_H^*|\cos(\ta1+\th0) =  
         \nonumber \\
  &&    \ \  -\frac{1}{\r2}v_1v_2\left\{|\lambda m_\phi^*|\cos(\ta2-\th0+\theta) + 
                  |m_3|\cos(\ta3+\th0+\theta)\right\},
  \label{extremum3}\\
  && |m_1^2|\sin\theta  = \frac{1}{\r2}v_0\left\{|\lambda \mu_\phi^*|\sin(\ta2-\th0+\theta)
                                                                              + |m_3|\sin(\ta3+\th0+\theta)\right\},
 \label{extremum4}\\
  && |m_2^2|v_0\sin2\th0  + \frac{1}{\r2}(v_1^2+v_2^2)|\lambda \mu_H^*|\sin(\ta1+\th0) =   
         \nonumber \\
  &&    \ \  -\frac{1}{\r2}v_1v_2\left\{|\lambda \mu_\phi^*|\sin(\ta2-\th0+\theta) 
                                    - |m_3|\sin(\ta3+\th0+\theta)\right\}.     
          \label{extremum5}                    
\end{eqnarray}
The assumption in Eq. (\ref{scale}) leads to the equations 
$\langle\partial V_1/\partial h_R^2\rangle =  \langle\partial V_1/\partial h_I^2\rangle = 0$, 
so that these extremum conditions are the same as those for the 
tree-level potential $V_0$.  

     The CP-even components and CP-odd ones of the Higgs scalar fields are mixed in 
the mass eigenstates.  
The mass-squared matrix for the neutral fields is expressed 
by a 6$\times$6 real symmetric matrix, denoted by $\cal M$.  
The elements of this matrix ${\cal M}_{ij}$ is given in Appendix.    
The matrix $\cal M$ is diagonalized by an orthogonal matrix $O$, 
\begin{equation}
   O^T{\cal M}O = {\rm diag}\left(\MH1, \MH2, \MH3, \MH4, \MH5, \MH6\right), 
\end{equation}
where the eigenvalues $\tilde M_{Hi}^2$ are in ascending order.  
The Higgs bosons in mass eigenstates $\Hi$ are then expressed as 
\begin{equation}
   \Hi = O_{1i}h_R^1 + O_{2i}h_R^2 + O_{3i}\phi_R + O_{4i}h_I^1 + O_{5i}h_I^2 + O_{6i}\phi_I.  
\end{equation}
The Goldstone boson for spontaneous breaking of SU(2) symmetry is represented 
by $\tilde H_1^0$ and thus the value of $\MH1$ vanishes.  

     Neglecting generation mixing for the quarks, the interaction Lagrangian for the Higgs bosons 
 and the $t$ or $b$ quark is given by 
\begin{eqnarray}
 \cal L &=& -\frac{m_t}{\r2 v_2}{\overline\psi_t}\left(F_u^i\PL + F_u^{i*}\PR\right)\psi_t\Hi 
      \nonumber \\
   & & -\frac{m_b}{\r2 v_1}{\overline\psi_b}\left(F_d^i\PL + F_d^{i*}\PR\right)\psi_b\Hi, 
       \label{int_tb}  \\
     F_u^i &=& {\rm e}^{-i\th2}(\O2i + i\O5i),  
     \quad 
       F_d^i = {\rm e}^{-i\th1}(\O1i + i\O4i),  
       \nonumber 
 \label{cpv}
 \end{eqnarray}
where $m_b$ denotes the $b$-quark mass.  
Even if generation mixing for the quarks is absent, CP invariance is not respected.  
It should be noted that the interactions of other $u$-type or $d$-type quarks are described by 
the same equations, 
provided that $m_t$ or $m_b$ is replaced with their masses.  
The interaction Lagrangian for the Higgs bosons and the $W$ or $Z$ boson is given by
\begin{eqnarray}
    {\cal L} &=& gM_W\sqrt{\frac{v_1^2+v_2^2}{v_1^2+v_2^2+4v_0^2}}G_W^iW^{+\mu}W_\mu^-\Hi + 
        \frac{1}{2}\sqrt{g^2+g'^2}M_Z G_Z^i Z^\mu Z_\mu\Hi, 
         \label{int_wz} \\
     G_W^i &=& \cos\beta(\O1i\cos\th1+\O4i\sin\th1) + \sin\beta(\O2i\cos\th2+\O5i\sin\th2) 
           \nonumber \\
                                  & & + \frac{4v_0}{\sqrt{v_1^2+v_2^2}}(\O3i\cos\th0+\O6i\sin\th0),                                
                                  \nonumber \\
      G_Z^i &=& \cos\beta(\O1i\cos\th1 +\O4i\sin\th1) +\sin\beta(\O2i\cos\th2+\O5i\sin\th2).    
      \nonumber       
\end{eqnarray}
The $Z$ boson does not couple to the SU(2)-triplet Higgs fields $\phi_R$ and $\phi_I$.   

\section{Two photon decay\label{decay}}

     The interactions of the Higgs bosons $\Hi$ ($i$=2-6) and the quarks could induce 
various CP-violating phenomena.  
In particular, sizable effects may be observed in the processes which quarks of the third generation 
participate in, since the coupling constants are generally proportional to the masses 
and thus non-negligible.  
One such process is the decay of the Higgs boson into two photons, 
which is generated dominantly through one-loop diagrams 
mediated by the $t$ or $b$ quark and $W$ boson \cite{resnick}.  
This decay has been studied for new particles which could mediate 
the process \cite{gunion}, though CP violation can also be probed.  

      In the two-photon decay, violation of CP invariance could be observed by measuring 
polarization of the photons.  
At the rest frame of the Higgs boson , the helicities of 
two photons are the same, both $h=+1$ or both $h=-1$.  
With $u(\pm,\vc p)$ denoting one photon state with helicity $\pm 1$ and momentum $\vc p$, 
the final state is written as $u(+,\vc p)u(+,-\vc p)$ or $u(-,\vc p)u(-,-\vc p)$.  
These two states are transformed to each other by CP operation, 
so that the eigenstates for CP-even and CP-odd are given respectively by 
\begin{eqnarray}
       f_{\rm even}&=&\frac{1}{\sqrt{2}}\left[u(+,\vc p)u(+,-\vc p)+u(-,\vc p)u(-,-\vc p)\right],  
       \label{cpeven}  \\
       f_{\rm odd}&=&\frac{1}{\sqrt{2}}\left[u(+,\vc p)u(+,-\vc p)-u(-,\vc p)u(-,-\vc p)\right].  
 \label{cpodd}
\end{eqnarray}
Since the Higgs boson is in mixed state of CP-even and CP-odd components, 
both of these final states appear, contrary to the SM Higgs boson.  
The two final states of CP-even and CP-odd could be distinguished from each other by  
observing photon polarization planes.  
In the CP-even state the polarization plane of one photon is parallel to that of another photon, 
while in the CP-odd state the two planes are perpendicular to each other.  
This difference can be detected by examining the angular distributions of 
the leptons or quarks which the photons internally convert to  \cite{kroll}, 
owing to their correlations with the polarization planes.  

     The decay widths for the CP eigenstates $ f_{\rm even}$ and  $f_{\rm odd}$ are  given by 
\begin{eqnarray}
   {\mathit\Gamma}_{\rm even} &=& \frac{e^4}{128\pi^5}\tilde M_{Hi}   \nonumber \\
    & & \left|\frac{2m_t}{3v_2}{\rm Re}(F_u^i)I(r_t) + \frac{m_b}{6v_1}{\rm Re}(F_d^i)I(r_b)  
                                       -\frac{g^2}{2M_W}\sqrt{v_1^2+v_2^2}\ G_W^iK(r_W)\right|^2, \\
   {\mathit\Gamma}_{\rm odd} &=& \frac{e^4}{128\pi^5}\tilde M_{Hi}
    \left|\frac{2m_t}{3v_2}{\rm Im}(F_u^i)J(r_t) + \frac{m_b}{6v_1}{\rm Im}(F_d^i)J(r_b)\right|^2, \\
    & & r_t=\frac{\tilde M_{Hi}}{m_t}, \quad  r_b=\frac{\tilde M_{Hi}}{m_b}, 
    \quad  r_W=\frac{\tilde M_{Hi}}{M_W}, 
\end{eqnarray}
where the functions are defined by 
\begin{eqnarray}
 I(r) &=& \frac{2}{r}\left[1-\left(\frac{4}{r^2}-1\right)f(r)\right], \\
 J(r) &=& \frac{2}{r}f(r), \\
 K(r) &=& \frac{r}{4} + \frac{3}{2r}\left[1-\left(\frac{4}{r^2}-2\right)f(r)\right],
\end{eqnarray}
\begin{eqnarray}
f(r) &=& \left(\arcsin\frac{r}{2}\right)^2  \quad\quad\quad\quad\quad\quad\quad\quad\quad (r\leq 2), \\
     &=& -\frac{1}{4}\left( \log\frac{r+\sqrt{r^2-4}}{r-\sqrt{r^2-4}}-i\pi\right)^2   \quad\quad (r>2).  
\end{eqnarray}
The QCD corrections to the quark contributions are small \cite{zerwas} and thus 
have been neglected.  

     The two-photon decay is also mediated by other particles, among which the charginos, 
the mixed states of charged Higgs fermions and SU(2) gauge fermions, could 
contribute sizably to both CP-even and CP-odd widths.  
However, these contributions depend additional model parameters, which makes the prediction 
less certain.  
Furthermore, non-negligible contributions are received only if the mass of the lighter chargino 
is of order of 100 GeV and thus accessible at LHC.  
In this study we do not incorporate the contributions mediated by the supersymmetric $R$-odd particles,  
assuming that these particles have large masses undetectable directly in the near future.    
The contributions from the charged Higgs bosons are small compared to the $W$ boson, 
and have also been neglected.  

\section{Numerical analyses\label{analyses}}

     The present model has various model parameters whose appropriate values are not 
known well.  
Instead of solving the extremum conditions in Eqs. (\ref{extremum1})-(\ref{extremum5}) for 
the VEVs, $i.e.$ $v_1$, $v_2$, $v_0$, $\theta$, and $\theta_0$, of the Higgs bosons, 
we express soft-breaking masses-squared ${\rm Re}(M_{H1}^2)$, ${\rm Re}(M_{H2}^2)$, 
${\rm Re}(M_\Phi^2)$, $|m_1^2|$, and $|m_2^2|$ in terms of the VEVs and the other 
parameters $|\lambda|$, $|\mu_H|$, $|\mu_\phi|$, $|m_3|$, and $\alpha_i$ ($i$=1-3).  
In case of $\sin\theta=0$ or $\sin 2\theta_0=0$, 
we take $|m_1^2|=|\mu_Hm_3|$ or $|m_2^2|=|\mu_\phi m_3|$, respectively.  
As for the magnitudes of the VEVs, the $Z$-boson mass gives the values of $v_1$ and $v_2$ 
from Eq. (\ref{Zmass}), with the ratio $v_2/v_1$ $(\equiv \tan\beta)$ being left undetermined.  
The value of $v_0$ is then constrained by Eq. (\ref{Wmass}).  
We fix $v_0$ at 3 GeV, which keeps the $W$ boson mass or $\rho$ parameter compatible with 
the experimental value \cite{Wexp}.
The remaining parameters are soft-breaking masses-squared Re($M_Q^2$) and Re($M_{U^c}^2$) 
in Eq. (\ref{stop}).  

     Given parameter values, the masses of the Higgs bosons and the coupling constants 
for the interactions in Eqs. (\ref{int_tb}) and (\ref{int_wz}) are determined.  
The lightest Higgs boson $\tilde H_2^0$ is considered to be the experimentally discovered particle.  
The mass has been known \cite{massexp} and  
the coupling constants are in rough agreement with the SM.  
The Higgs boson is produced dominantly through the gluon fusion mediated 
by the $t$ quark.  
Then, the production cross section $\sigH2$ is proportional 
roughly to the square of the coupling constant for $t$ and $\bar t$.  
The decay widths for $\bar bb$, $WW^*$, and $ZZ^*$ are also proportional to the square 
of their coupling constants.  
Since the Higgs boson decays dominantly into $b$ and $\bar b$, 
the branching ratios Br($WW^*$) and Br($ZZ^*$) may be given by the 
ratios of their widths to the width for $\bar bb$.  
Therefore, the ratios of this model to the SM for the cross section and branching ratios 
could be estimated roughly by the ratios of the coupling constants.  

     Allowing for uncertainty of our scheme and experimental results, we impose the 
following constraints on the parameters,  
\begin{eqnarray}
  &&\tilde M_{H2} = 121-131 \ \ [\rm GeV], 
  \\
  &&\frac{\sigH2}{\sigSMH2} \simeq \frac{v_1^2+v_2^2+4v_0^2}{v_2^2}|F_u^2|^2 = 0.6 - 1.4, 
  \\
  &&\frac{\sigH2\cdot{\rm Br}(WW^*)}{\sigSMH2\cdot{\rm Br}_{SM}(WW^*)} \simeq  
  \cot^2\beta\frac{v_1^2+v_2^2}{v_1^2+v_2^2+4v_0^2}\frac{|F_u^2|^2}{|F_d^2|^2}(G_W^2)^2 
      = 0.6 - 1.4,
  \\
   &&\frac{\sigH2\cdot{\rm Br}(ZZ^*)}{\sigSMH2\cdot{\rm Br}_{SM}(ZZ^*)} \simeq 
              \cot^2\beta\frac{|F_u^2|^2}{|F_d^2|^2}(G_Z^2)^ 2= 0.6 - 1.4.  
\end{eqnarray}
Here, $\sigma_{SM}$ and ${\rm Br}_{SM}$ denote the cross section and the branching ratio 
under the SM interactions, with $\tilde M_{H2}$ being taken for the Higgs boson mass.  
The decay width for $\bar bb$ receives non-negligible contributions from 
QCD corrections \cite{kniehl}.  
However, the above estimate uses the ratio for the two models, which is not affected much 
by the corrections.  

     Another experimental constraint could come from non-observation of a Higgs-like boson, 
other than the observed one, for the mass range smaller than 710 GeV \cite{higgs3}.  
In the present model there exist four extra neutral Higgs bosons.  
However, these Higgs bosons are predicted to show phenomena different from the SM Higgs boson.  
For instance, in wide region of parameter space the second lightest Higgs boson $\tilde H_3^0$ 
has a large branching ratio for the decay into $\bar bb$.  
On the other hand, the Higgs boson in the SM would decay into $W^+W^-$ and $ZZ$ dominantly, 
if the mass is larger than twice the $W$ boson mass. 
In fact, these decay modes have been explored in experiments for an SM Higgs boson 
with a large mass.   
Therefore, we do not further constrain the parameters by the extra Higgs bosons.  

     The experimental constraints on the Higgs boson $\tilde H_2^0$ are satisfied 
in wide region of parameter space.  
In the following numerical calculations we take two sets of values for $|\mu_H|$ and $|\mu_\phi|$, 
$(a)$ $|\mu_H|=300$ GeV, $|\mu_\phi|=300$ GeV and 
$(b)$ $|\mu_H|=1000$ GeV, $|\mu_\phi|=1000$ GeV, with $|\lambda|=1$, $|m_3|=1000$ GeV, 
Re($M_Q^2$)= Re($M_{U^c}^2$)=1000 GeV.  
Two complex phases are fixed as $\alpha_1=\pi/2$ and $\alpha_2=-\pi/4$, and the other 
phases $\alpha_3$, $\theta_0$, and $\theta$ are kept variable.  

\begin{figure}[htbp]
\begin{center}
\begin{tabular}{c}

\begin{minipage}{0.55\hsize}
\begin{center}
\includegraphics[width=8 cm]{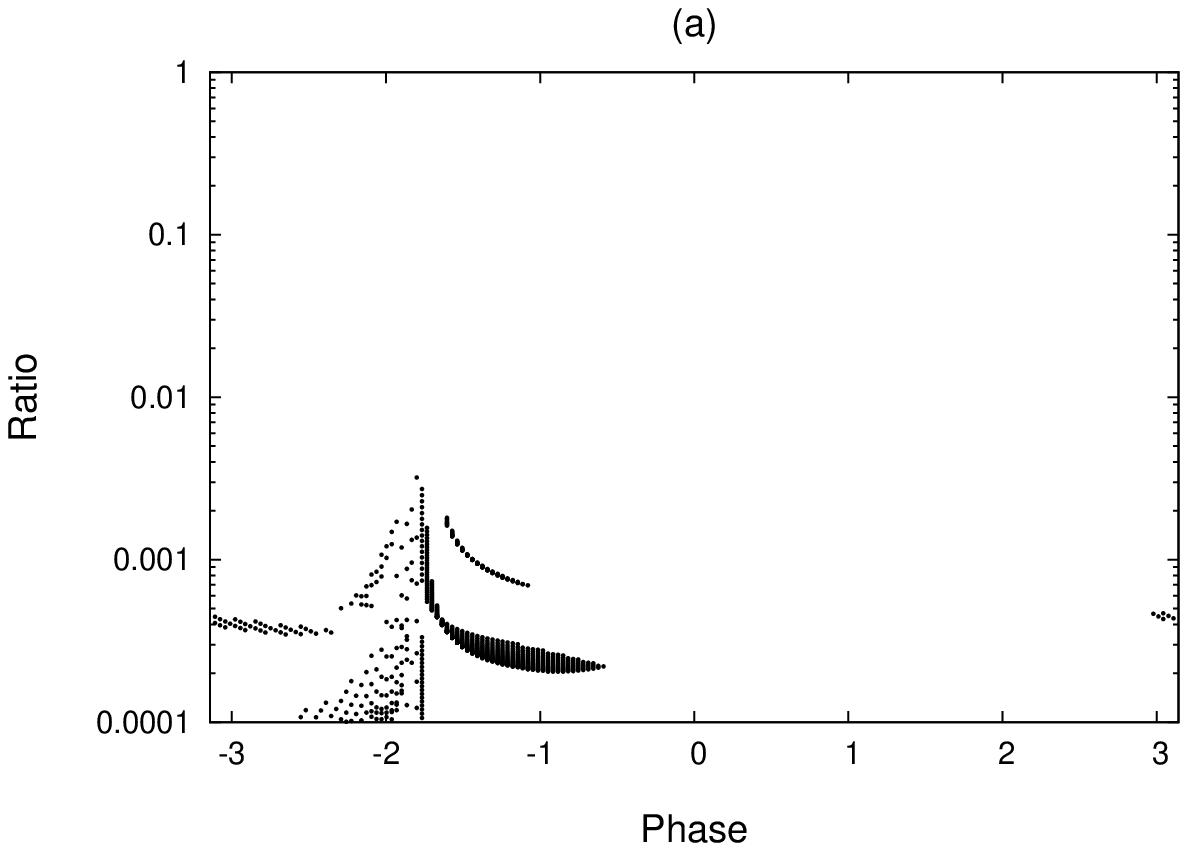} \\
\end{center}
\end{minipage}

\begin{minipage}{0.55\hsize}
\begin{center}
\includegraphics[width=8 cm]{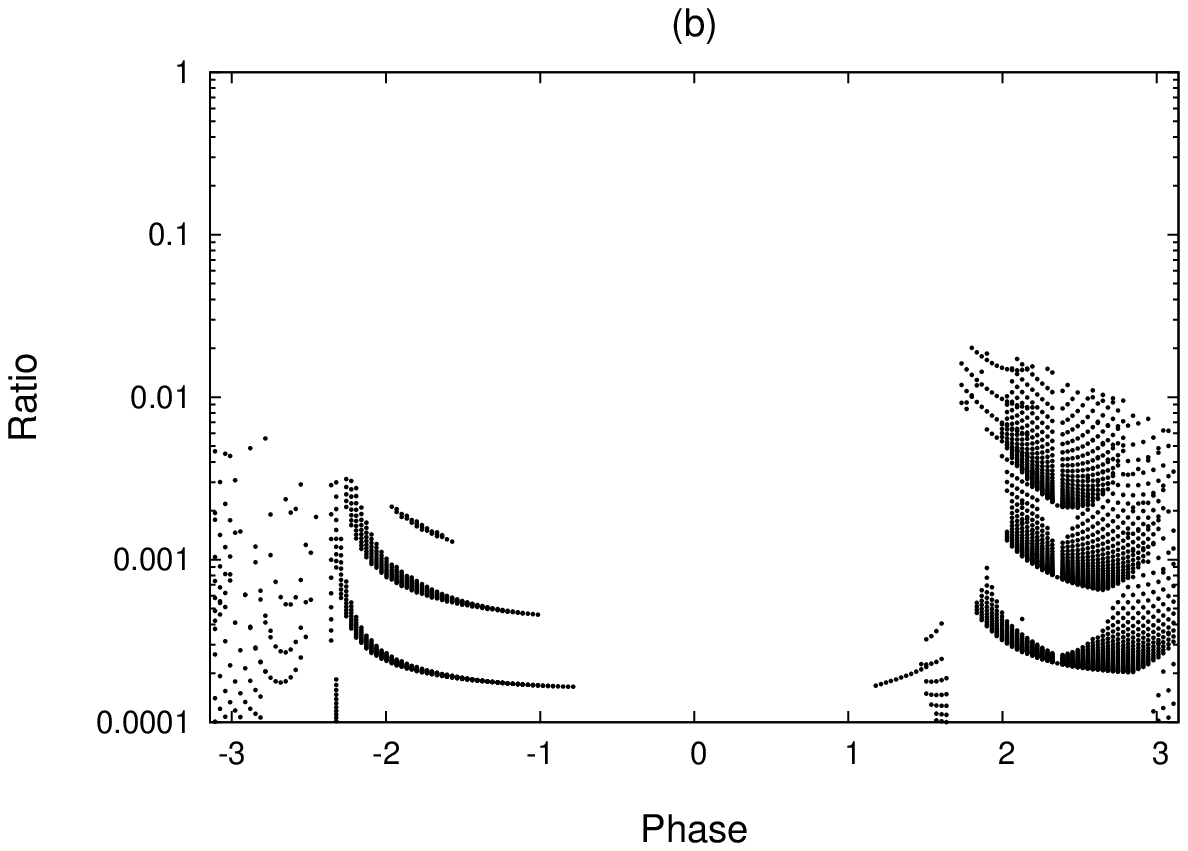} \\
\end{center}
\end{minipage}

\end{tabular}
\vspace{1 cm}
\caption{The ratio ${\mathit\Gamma}_{\rm odd}/{\mathit\Gamma}_{\rm even}$ 
of the decay $\tilde H_2^0\to\gamma\gamma$ for $\tan\beta=2$.  
The horizontal axis stands for the phase $\alpha_3$.  
$(a)$ $|\mu_H|=300$ GeV, $|\mu_\phi|=300$ GeV;  
$(b)$ $|\mu_H|=1000$ GeV, $|\mu_\phi|=1000$ GeV.}
\label{ratio2}
\end{center}
\end{figure}

     We first show the CP-violating effect on the two-photon decay of the lightest Higgs boson $\tilde H_2^0$.  
In Fig. \ref{ratio2} the ratio $R$ of ${\mathit\Gamma}_{\rm odd}$ to 
${\mathit\Gamma}_{\rm even}$ is depicted as a function of $\alpha_3$ for $\tanb=2$.  
With $\theta_0$ and $\theta$ being varied, if the vacuum is consistent with the experimental constraints, 
the ratio $R$ is indicated as a dot.  
It is seen that the two-photon decay could yield both CP-odd and CP-even final states 
at a ratio of order of $10^{-2}-10^{-3}$.  
As the value of $\tanb$ becomes large, the ratio decreases; $R<10^{-4}$ for $\tanb\sim10$.  
The dominant contribution to the decay is mediated by 
the $W$ boson which does not yield the CP-odd final state, so that the magnitude of $R$ is not large.  
For some parameter values, however, both CP-even and 
CP-odd components are contained sizably in the Higgs boson, as shown in Table \ref{comptab}.  
The branching ratio of $\tilde H_2^0 \to \gamma\gamma$ is around $2\times 10^{-3}$, which is 
not different much from the SM and compatible with the experimental results.  
If the Higgs boson is produced at the number of $10^6-10^7$, the appearance of CP-odd polarization  
state may be detectable.  

\begin{table}
\caption{
The components of the lightest Higgs boson for $\tanb=2$.  \\
$(a)$ $\alpha_1=\pi/2$, $\alpha_2=-\pi/4$, $\alpha_3=-\pi/2$, $\theta_0=-3\pi/4$, $\theta=\pi/48$, 
$(b)$ $\alpha_1=\pi/2$, $\alpha_2=-\pi/4$, $\alpha_3=7\pi/8$, $\theta_0=-\pi/4$, $\theta=\pi/24$.
\label{comptab}
}
\vspace{1 cm}
\begin{ruledtabular}
\begin{tabular}{ccccccc}
  & $h_R^1$ & $h_R^2$ & $\phi_R$  & $h_I^1$ & $h_I^2$ & $\phi_I$ \\
\hline
$(a)$ & 0.48 & 0.87 & $-0.37\times 10^{-1}$ & 0.11 
                                                      & 0.40$\times 10^{-1}$ & $-0.25\times 10^{-1}$\\
$(b)$ & $-0.33$ & $-0.87$ & -0.15$\times 10^{-2}$ & $-0.34$
                                                      & $-0.14$ & 0.35$\times 10^{-1}$\\
\end{tabular}
\end{ruledtabular}
\end{table}

\begin{figure}[htbp]
\begin{center}
\begin{tabular}{c}

\begin{minipage}{0.55\hsize}
\begin{center}
\includegraphics[width=8 cm]{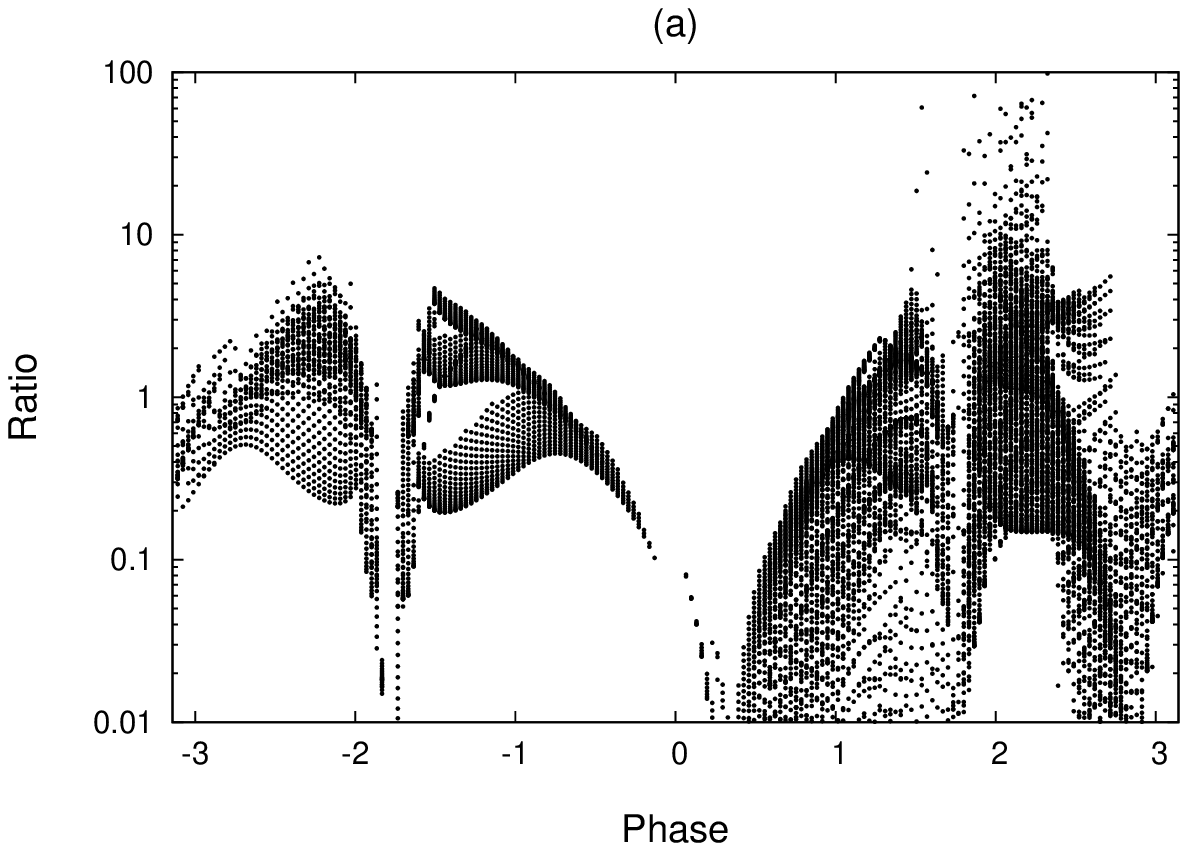} \\
\end{center}
\end{minipage}

\begin{minipage}{0.55\hsize}
\begin{center}
\includegraphics[width=8 cm]{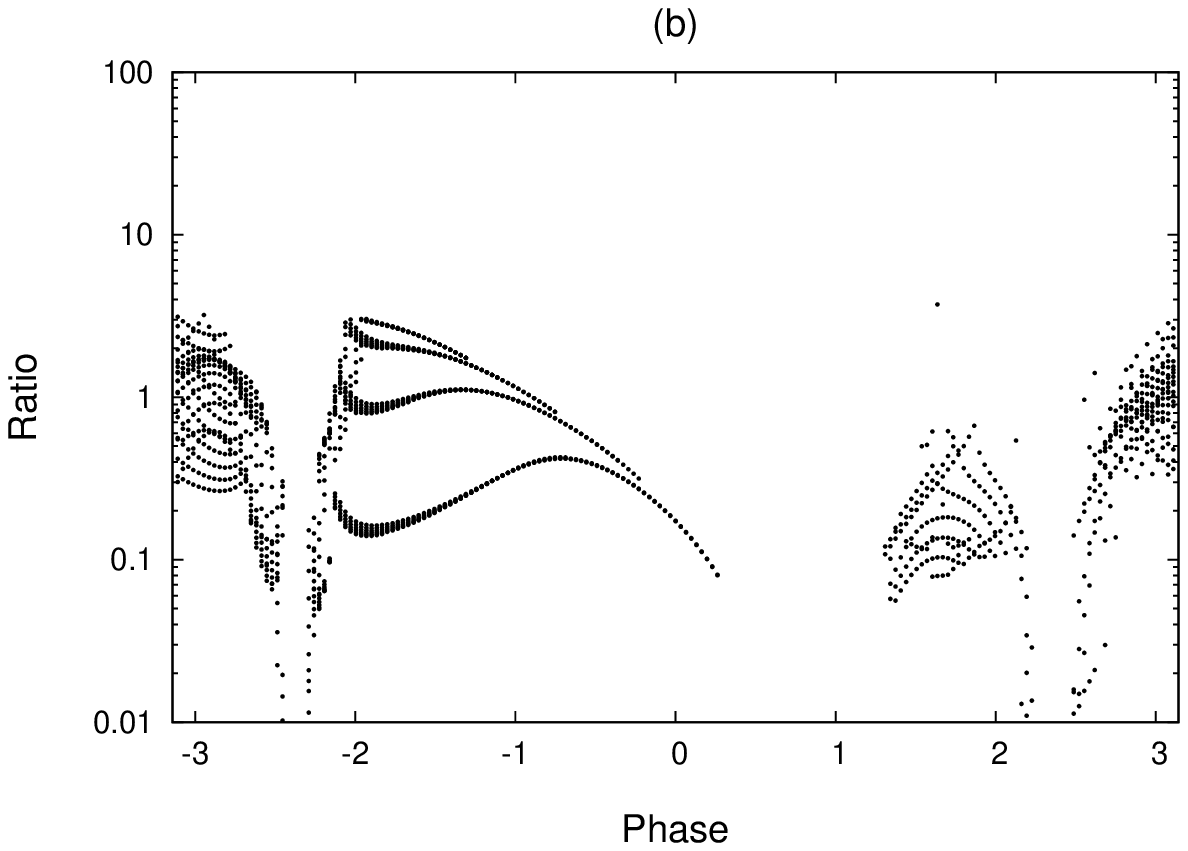} \\
\end{center}
\end{minipage}

\end{tabular}
\vspace{1 cm}
\caption{The ratio ${\mathit\Gamma}_{\rm odd}/{\mathit\Gamma}_{\rm even}$  
of the decay $\tilde H_3^0\to\gamma\gamma$ for $\tan\beta=10$.  
The horizontal axis stands for the phase $\alpha_3$.  
$(a)$ $|\mu_H|=300$ GeV, $|\mu_\phi|=300$ GeV;  
$(b)$ $|\mu_H|=1000$ GeV, $|\mu_\phi|=1000$ GeV.}
\label{ratio10}
\end{center}
\end{figure}

\begin{figure}[htbp]
\begin{center}
\begin{tabular}{c}

\begin{minipage}{0.55\hsize}
\begin{center}
\includegraphics[width=8 cm]{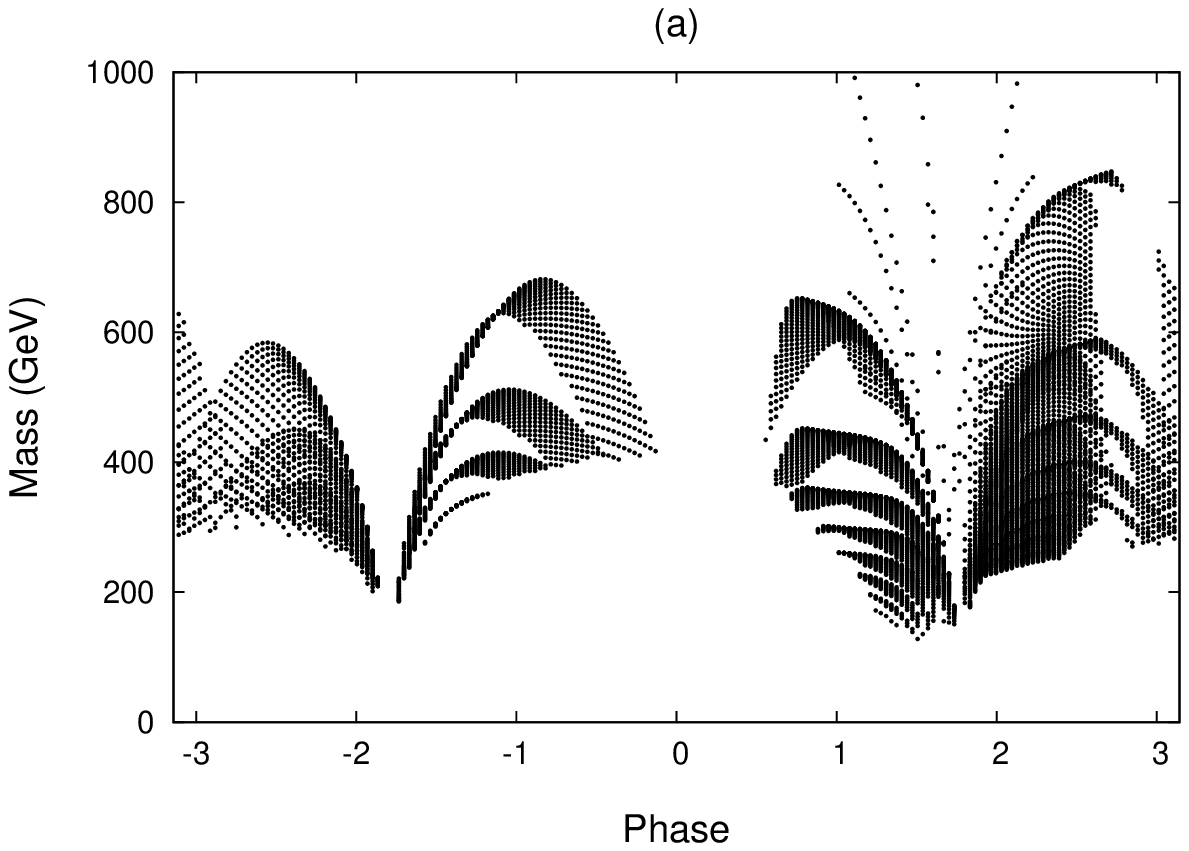} \\
\end{center}
\end{minipage}

\begin{minipage}{0.55\hsize}
\begin{center}
\includegraphics[width=8 cm]{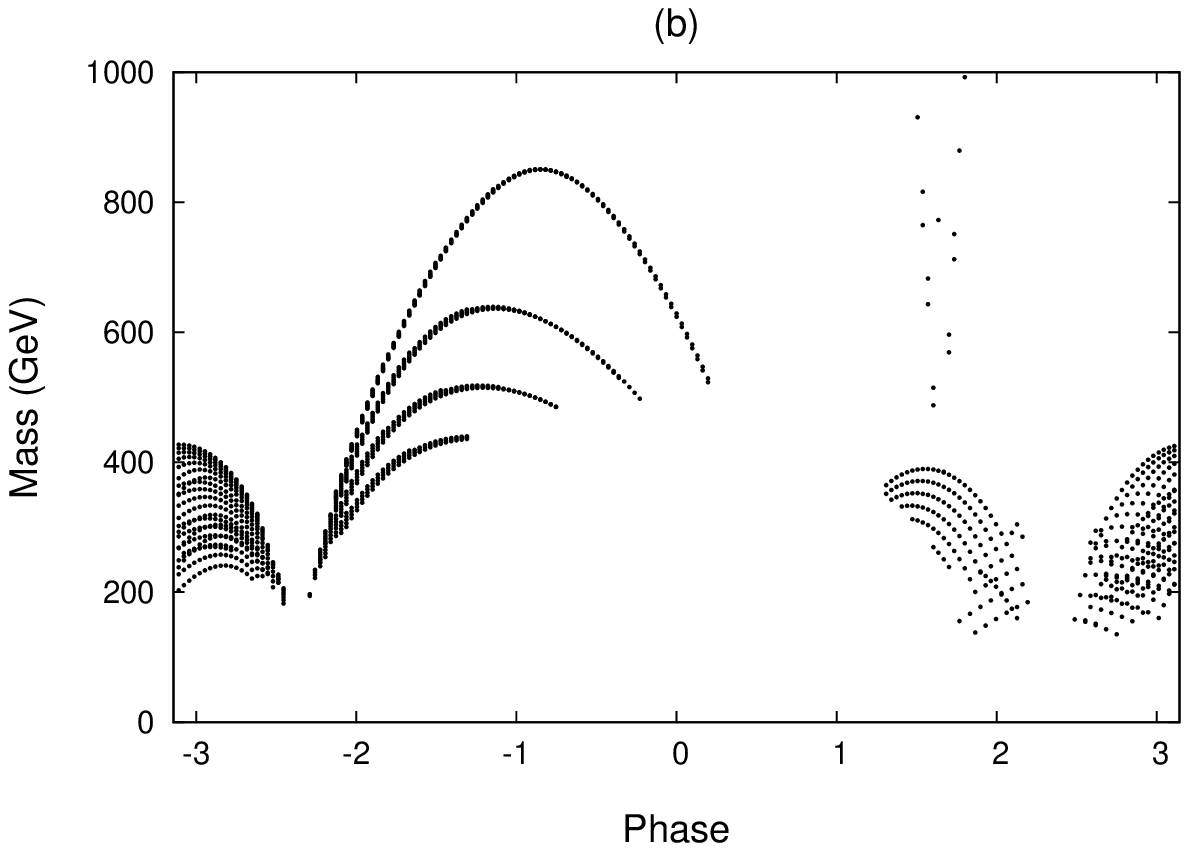} \\
\end{center}
\end{minipage}

\end{tabular}
\vspace{1 cm}
\caption{The mass of the second lightest Higgs boson $\tilde H_3^0$ for $\tan\beta=10$ 
 under the constraint $0.1<{\mathit\Gamma}_{\rm odd}/{\mathit\Gamma}_{\rm even}<10.0$.}
\label{mass}
\end{center}
\end{figure}

\begin{figure}[htbp]
\begin{center}
\begin{tabular}{c}

\begin{minipage}{0.55\hsize}
\begin{center}
\includegraphics[width=8 cm]{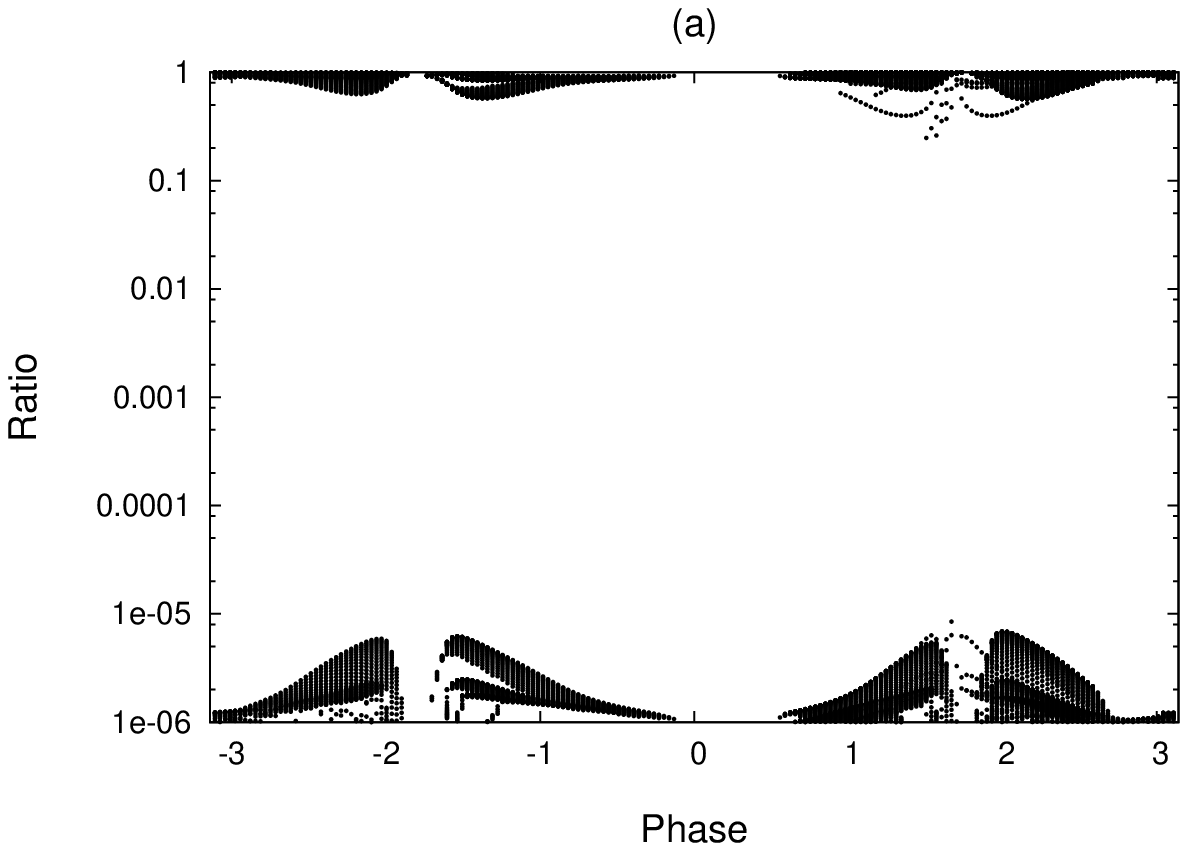} \\
\end{center}
\end{minipage}

\begin{minipage}{0.55\hsize}
\begin{center}
\includegraphics[width=8 cm]{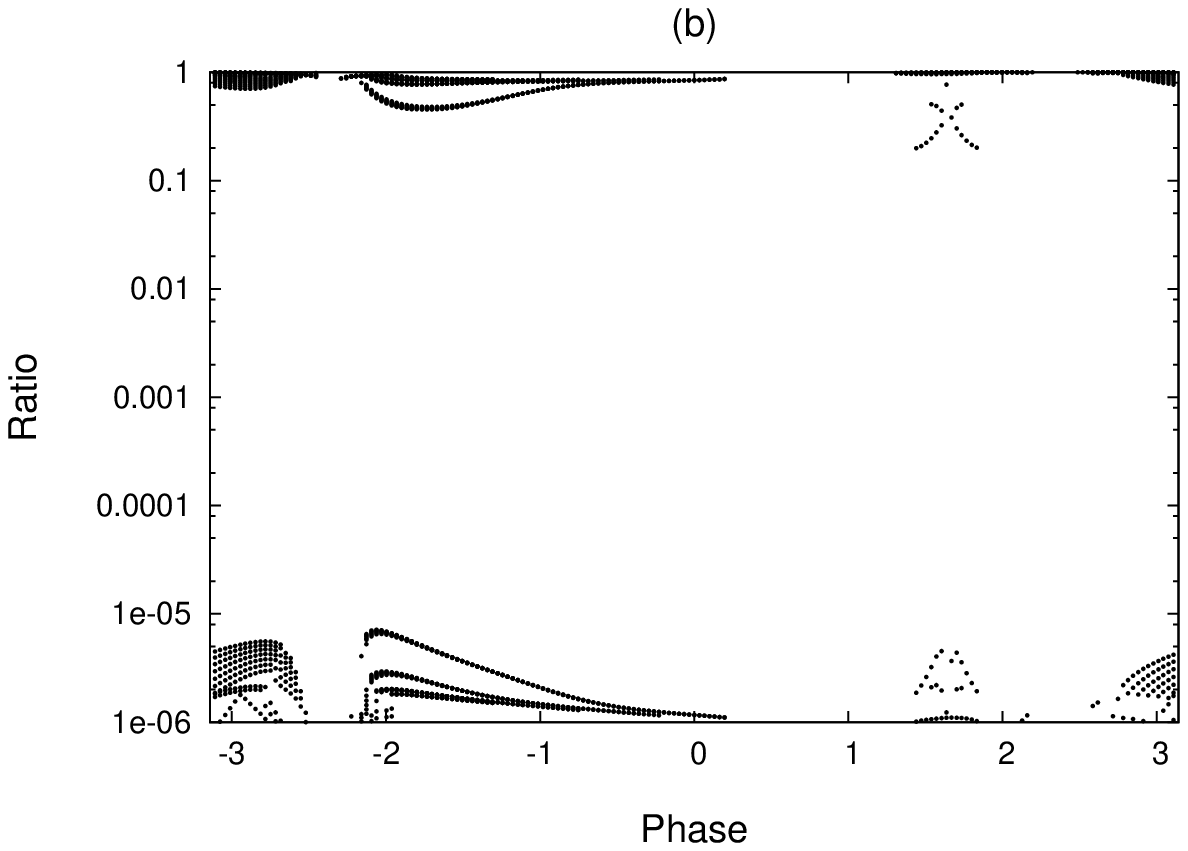} \\
\end{center}
\end{minipage}

\end{tabular}
\vspace{1 cm}
\caption{The branching ratios of the second lightest Higgs boson for $\tan\beta=10$ 
under the constraint $0.1<{\mathit\Gamma}_{\rm odd}/{\mathit\Gamma}_{\rm even}<10.0$.  
The upper and lower dots correspond respectively to $\tilde H_3^0\to \bar bb$ and 
$\tilde H_3^0\to\gamma\gamma$.}
\label{bratio}
\end{center}
\end{figure}

     Comparable rates for CP-even and CP-odd final states could be observed in 
the two-photon decay of the second lightest Higgs boson $\tilde H_3^0$.  
In Fig. \ref{ratio10} the ratio $R$ of ${\mathit\Gamma}_{\rm odd}$ to 
${\mathit\Gamma}_{\rm even}$ is depicted for $\tanb=10$.  
In wide region of parameter space the ratio becomes of order of unity, 
which holds also for smaller values of $\tanb$.  
The mass of $\tilde H_3^0$ is shown in Fig. \ref{mass}, 
where the parameter values are further constrained to satisfy $0.1<R<10.0$.    
The Higgs boson with the ratio $R\sim 1$ could be rather light enough for 
detection at the LHC.  
However, its phenomenological nature is different much from the Higgs-like boson 
of the SM.  
In Fig. \ref{bratio} the branching ratio of $\tilde H_3^0 \to \bar b b$ is shown, 
together with that of $\tilde H_3^0 \to \gamma\gamma$.   
The decay to $\bar bb$ has the largest branching ratio, while in the SM the Higgs-like boson 
with a mass larger than about 200 GeV decays dominantly into $W^+W^-$ and $ZZ$.  
As explicit numerical examples, we give in Table \ref{bratiotab} the branching ratios 
for two sets of parameter values.  
In order to detect the Higgs boson, it would be necessary to make experimental analyses 
which are different from those for searching a Higgs-like boson of the SM.  
Although the branching ratio of the two-photon decay is small, both CP-odd and CP-even final states 
could appear at the rates of the same order of magnitude.  
The number of $10^6-10^7$ for the Higgs boson could enable detection of CP violation.  

\section{Discussions\label{discussions}}
 
      We have studied the Higgs bosons in the supersymmetric model 
which has an extra Higgs superfield of SU(2)-triplet representation.  
The Higgs sector then induces naturally violation of CP invariance at tree level, 
which does not occur in the SM nor in the MSSM.  
Although the experimental results for the observed Higgs boson 
constrain extensions of the Higgs sector, 
it has been shown that  there is still room for our CP-violating potential.  
Any phenomena which do not conserve CP symmetry for the Higgs bosons 
would provide us an important clue for physics beyond the SM.  

\begin{table}
\caption{
The mass and branching ratios of the second lightest Higgs boson for $\tanb=10$.  \\
$(a)$ $\alpha_1=\pi/2$, $\alpha_2=-\pi/4$, $\alpha_3=5\pi/8$, $\theta_0=-\pi/4$, $\theta=\pi/48$, 
$(b)$ $\alpha_1=\pi/2$, $\alpha_2=-\pi/4$, $\alpha_3=-5\pi/8$, $\theta_0=-23\pi/24$, $\theta=\pi/48$.
\label{bratiotab}
}
\vspace{1 cm}
\begin{ruledtabular}
\begin{tabular}{ccccccc}
  & mass (GeV) & $\bar bb$ & $\bar tt$  & $W^+W^-$ & $ZZ$ & $\gamma\gamma$ \\
\hline
$(a)$ & 0.40$\times 10^3$ & 0.89& 0.83$\times 10^{-1}$ & 0.36$\times 10^{-2}$ 
                                                      & 0.20$\times 10^{-1}$ & 0.21$\times 10^{-5}$\\
$(b)$ & 0.44$\times 10^3$ & 0.79 & 0.14 & 0.67$\times 10^{-2}$ 
                                                      & 0.67$\times 10^{-1}$ & 0.29$\times 10^{-5}$\\
\end{tabular}
\end{ruledtabular}

\
\end{table}

     One of the CP-violating effects of the Higgs sector may be observed in the polarization 
of the two photons coming from the Higgs boson decay.  
The observed Higgs boson, which is consistent with the SM and should then be CP-even, 
could yield a CP-odd final state at the ratio of $10^{-3}-10^{-2}$.  
The second lightest Higgs boson could give both CP-odd and CP-even states
at comparable rates.  
Although existence of an SM-like Higgs boson has been ruled out in a wide mass range, 
the extra Higgs boson has decay property different much from the SM and thus could 
have escaped detection.  
Searching for a new Higgs boson from various aspects is awaited.  

      Coexistence of both CP eigenstates for the two-photon decay could also be found in the 
framework of the MSSM, though the Higgs bosons are in either of the CP eigenstates.  
This non-conservation is realized by the charginos whose interactions with the Higgs bosons 
do not respect CP invariance.  
However, the mixing of CP eigenstates becomes non-negligible only through the chargino
of oder of 100 GeV, which would be detectable in near future experiments.   
In case all the supersymmetric $R$-odd particles are sufficiently heavy, the MSSM 
do not predict CP violation in the Higgs sector, similarly to the SM.  

\appendix*\section{} 

     The mass-squared matrix $\cal M$ for the neutral Higgs bosons 
is expressed as the sum of the matrices ${\cal M}^0$ and ${\cal M}^1$ coming respectively from 
the tree-level potential and the one-loop potential.    
The elements for ${\cal M}^0$ are given by 
\begin{eqnarray}
\M11 &=& M_1^2 + 2r_1(1+2\cos^2\th1)v_1^2 + r_2v_2^2+r_3v_0^2 + 
                         \r2|\lambda\mu_H^*|v_0\cos(\ta1+\th0), 
    \\
\M12 &=& 2r_2v_1v_2\cos\th1\cos\th2 - |m_1^2 | 
      \nonumber   \\
         & & +\frac{1}{\r2}v_0\{|\lambda\mu_\phi^*| \cos(\ta2-\th0) + |m_3|\cos(\ta3+\th0)\},
   \\
\M13 &=& 2r_3v_1v_0\cos\th1\cos\th0+ \r2 v_1 |\lambda\mu_H^*|\cos\ta1\cos\th1 
    \nonumber   \\
        & & +\frac{1}{\r2}v_2\{|\lambda\mu_\phi^*|\cos(\ta2+\th2) + |m_3|\cos(\ta3+\th2)\},
   \\
\M14 &=& 2r_1v_1^2\sin2\th1, 
 \\
\M15 &=& 2r_2v_1v_2\cos\th1\sin\th2 - \frac{1}{\r2}v_0\{|\lambda\mu_\phi^*|\sin(\ta2-\th0) + 
                     |m_3|\sin(\ta3+\th0)\}, 
  \\
\M16 &=& 2r_3v_1v_0\cos\th1\sin\th0 + \r2 v_1|\lambda\mu_H^*|\sin\ta1\cos\th1  
    \nonumber   \\
       & & + \frac{1}{\r2}v_2\{|\lambda\mu_\phi^*|\sin(\ta2+\th2) - |m_3|\sin(\ta3+\th2)\}, 
  \\
\M22 &=& M_2^2 + 2r_1v_2^2(1+2\cos^2\th2) + r_2v_1^2+r_3v_0^2 - 
                      \r2v_0|\lambda\mu_H^*|\cos(\ta1+\th0),
       \\
\M23 &=& 2r_3v_2v_0\cos\th2\cos\th0 - \r2 v_2|\lambda\mu_H^*|\cos\ta1\cos\th2
    \nonumber   \\
       & & + \frac{1}{\r2}v_1\{|\lambda\mu_\phi^*|\cos(\ta2+\th1) + |m_3|\cos(\ta3+\th1)\}, 
  \\
\M24 &=& 2r_2v_1v_2\sin\th1\cos\th2 
    \nonumber   \\
       & & - \frac{1}{\r2}v_0\{|\lambda\mu_\phi^*|\sin(\ta2-\th0) + |m_3|\sin(\ta3+\th0)\}, 
  \\
\M25 &=& 2r_1v_2^2\sin2\th2, 
 \\
\M26 &=& 2r_3v_2v_0\cos\th2\sin\th0 + \r2 v_2|\lambda\mu_H^*|\sin\ta1\cos\th2 
    \nonumber   \\
       & & + \frac{1}{\r2}v_1\{|\lambda\mu_\phi^*|\sin(\ta2+\th1) - |m_3|\sin(\ta3+\th1)\}, 
  \\
\M33 &=& M_3^2 + r_3(v_1^2+v_2^2) - |m_2^2|
      \\
\M34 &=& 2r_3v_1v_0\sin\th1\cos\th0 - \r2 v_1|\lambda\mu_H^*|\cos\ta1\sin\th1  
   \nonumber    \\
       & & - \frac{1}{\r2}v_2\{|\lambda\mu_\phi^*|\sin(\ta2+\th2) + |m_3|\sin(\ta3+\th2)\}, 
  \\
\M35 &=& 2r_3v_2v_0\sin\th2\cos\th0 - \r2 v_2|\lambda\mu_H^*|\cos\ta1\sin\th2 
    \nonumber   \\
       & & - \frac{1}{\r2}v_1\{|\lambda\mu_\phi^*|\sin(\ta2+\th1) + |m_3|\sin(\ta3+\th1)\}, 
  \\
\M36 &=& 0,
  \\
\M44 &=& M_1^2 + 2r_1v_1^2(1+2\sin^2\th1) + r_2v_2^2+r_3v_0^2 - 
                    \r2v_0|\lambda\mu_H^*|\cos(\ta1+\th0),
       \\
\M45 &=& 2r_2v_1v_2\sin\th1\sin\th2 + |m_1^2|  
   \nonumber   \\
         & & -\frac{1}{\r2}v_0\{|\lambda\mu_\phi^*|\cos(\ta2-\th0) + |m_3|\cos(\ta3+\th0)\},
   \\
\M46 &=& 2r_3v_1v_0\sin\th1\sin\th0 + \r2 v_1|\lambda\mu_H^*|\sin\ta1\sin\th1  
    \nonumber   \\
       & & + \frac{1}{\r2}v_2\{|\lambda\mu_\phi^*|\cos(\ta2+\th2) - |m_3|\cos(\ta3+\th2)\}, 
  \\
\M55 &=& M_2^2 + 2r_1v_2^2(1+2\sin^2\th2) + r_2v_1^2+r_3v_0^2 - 
                  \r2 v_0|\lambda\mu_H^*|\cos(\ta1+\th0),
       \\
\M56 &=& 2r_3v_2v_0\sin\th2\sin\th0 + \r2 v_2|\lambda\mu_H^*|\sin\ta1\sin\th2  
   \nonumber    \\
       & & + \frac{1}{\r2}v_1\{|\lambda\mu_\phi^*|\cos(\ta2+\th1) - |m_3|\cos(\ta3+\th1)\}, 
  \\
\M66 &=& M_3^2 + r_3(v_1^2+v_2^2) + |m_2^2|, 
\end{eqnarray}
where the extremum conditions in Eqs. (\ref{extremum1})-(\ref{extremum5}) are not taken into account.  
The elements for ${\cal M}^1$ are given by 
\begin{eqnarray}
{\cal M}_{22}^1 &=& \frac{3}{8\pi^2}\frac{m_t^4}{v_2^2}\cos^2\th2\log\frac{M_{t1}^2M_{t2}^2}{m_t^4}, 
\\
{\cal M}_{55}^1 &=& \frac{3}{8\pi^2}\frac{m_t^4}{v_2^2}\sin^2\th2\log\frac{M_{t1}^2M_{t2}^2}{m_t^4}, 
\\
{\cal M}_{25}^1 &=& 
                  \frac{3}{8\pi^2}\frac{m_t^4}{v_2^2}\sin\th2\cos\th2\log\frac{M_{t1}^2M_{t2}^2}{m_t^4}.  
\end{eqnarray}
The indices $i, j$ (=1-6) are 
in order of $(h_R^1,h_R^2, \phi_R, h_I^1, h_I^2, \phi_I)$.


\end{document}